# 3D landmark detection for augmented reality based otologic procedures

**Raabid HUSSAIN, Caroline GUIGOU, Kibrom Berihu GIRUM, Alain LALANDE and Alexis BOZORG GRAYELI**

*ImViA Laboratory, Université de Bourgogne Franche-Comté, 21000 Dijon, France*

*Tel : +33 (0)6 95 96 55 40*

*Contact: raabid.hussain@u-bourgogne.fr*

**E**ar consists of the smallest bones in the human body and does not contain significant amount of distinct landmark points that may be used to register a preoperative CT-scan with the surgical video in an augmented reality framework. Learning based algorithms may be used to help the surgeons to identify landmark points. This paper presents a convolutional neural network approach to landmark detection in preoperative ear CT images and then discusses an augmented reality system that can be used to visualize the cochlear axis on an otologic surgical video.

## 1 Introduction

Augmented reality (AR) has been widely accepted and applied in different surgical domains like orthopedics, hepatobiliary and pancreatic systems. However, owing to minuscule operative space, limited field of view and complex instrument trajectory, AR has not been widely accepted for otologic procedures. Otologic structures constitute some of the smallest structures in the human body and their handlings require submillimetric precision and expert knowledge of intra and inter anatomic relations as critical nerves and vessels are in close proximity.

Image registration is one of the critical processes in an AR system. In otology, preoperative computed tomography (CT) is particularly helpful as the ear is composed of mostly rigid bony structures. Precise identification of anatomical landmarks for registration is often difficult and time consuming as the points are not well defined. Due to low number of prominent features between preoperative CT reconstructions and endoscopic videos, different studies have used fiducial markers to register the pre and intra-operative images [1, 2].

This study aims at simultaneously determining locations of multiple anatomical landmarks in preoperative CT-scan using a learning based approach. The detected landmarks may then be used to register CT with microscopic images. Typical learning models employ Gaussian heat maps to optimally determine the landmark position [3]. However, since the distance between each landmark is very small, this is not a viable option in this scenario. Moreover, state-of-the-art learning architectures often require manual fine-tuning of meta-parameters which often leads to loss of important information. This study proposes and evaluates the performance of a convolutional neural network (CNN) for middle and inner ear landmark detection without standardizing the meta-parameters and tests them on an AR scenario to infer the cochlear axis.

## 2 Methodology

### 2.1 Dataset

We used a dataset of 25 patients (age range: 15-77 years) comprising of 40 ear CTs (17 right and 23 left ears) acquired from different scanners. The pixel resolutions of the acquired CT scans ranged from $0.156 \times 0.156 \times 0.100$ mm$^3$ to $0.292 \times 0.292 \times 0.625$ mm$^3$. The x-y image size was the same for all the images: 512 x 512 while the z image size ranged from 92 to 876. Locations of seven anatomical landmarks were identified in all the scans by an expert surgeon:

round window niche, tip of the incus, umbo and short process of malleus, pyramid, cochlear apex and base.

## 2.2 Landmark detection

To homogenize the data, the left ear CTs were flipped to resemble the right ear CTs. Due to memory limitations, the surgeon was asked to crop a 200x200x100 region of interest from the original CT data that contained the middle and inner ear contents. The region of interest was passed through a convolutional neural network depicted in Figure 1. All layers had 'elu' activations except the final layer which had linear activation. The output layer had 21 units comprising of x, y and z coordinates of each landmark. The training was carried out for 3500 epochs with a batch size of 5, using Adam optimizer with a learning rate of 0.0005, and mean squared logarithmic error as the loss function. The architecture was implemented on a computer with dedicated GPU (NVIDIA TITAN X, 12 GB RAM processor) using Keras and Tensorflow libraries. The network was trained from scratch and assessed using 5 fold cross validation approach.

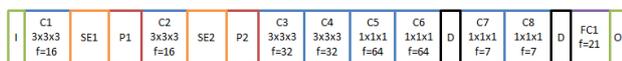

*Figure 1: CNN for landmark detection. I=Input layer, C=Convolutional layer, SE=Squeeze and excitation block [4], P=Max pooling layer, FC=Fully-connected layer, D=Dropout with a rate of 0.2, O= Output layer, f=number of filters.*

## 2.3 AR system

An AR system was developed to test the output of the landmark detection step. Cochlear base point was used as a test landmark as it is obscured in microscopic view. Remaining six landmarks were used to register the preoperative CT with the real-time microscopic video. The surgeon was asked to select their corresponding points in the initial microscope image. Fundamental matrix of the microscope camera (determined through the 2D points in microscope image and their respective 3D coordinates in CT image) was used to register the two information, followed by the use of a SURF based feature matching to track the motion of the microscope (similar system as used in [5]). The cochlear axis (defined by cochlear apex and base points) was drawn onto the microscopic video for the surgeon to have more insight about the inner ear. The system was tested on a phantom resin model of the temporal bone with the tympanic membrane removed.

## 3 Results

The landmark detection on CT-scan yielded an average error of 0.88 ± 0.27 mm (mean absolute distance between position of the detected landmark and its location as indicated by surgeon). Individual errors for each landmark are listed in Table 1.

*Table 1: Landmark detection process accuracy. R=Round window niche, I=Incus tip, U=Umbo of malleus, S=Short process of malleus, P=Pyramid tip, A=Cochlear apex, B=Cochlear base*

|               | R    | I    | U    | S    | P    | A    | B    |
|---------------|------|------|------|------|------|------|------|
| Error (mm)    | 0.82 ± 0.36 | 0.71 ± 0.22 | 0.88 ± 0.26 | 0.92 ± 0.31 | 0.89 ± 0.27 | 0.93 ± 0.16 | 1.05 ± 0.31 |

The AR system is displayed in Figure 2. An ENT expert verified that the position of the cochlear axis was indeed in close proximity to where the axis was expected. The system maintained correspondence throughout the experimental time of 2 minutes.

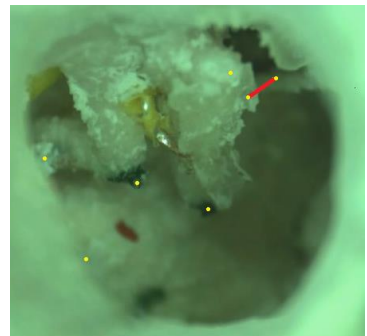

*Figure 2: Augmented reality display. The cochlear axis is shown with a red line and the landmark points are represented by yellow dots.*

## 4 Conclusion

This paper presented a CNN based otologic landmark detection and its potential application in AR based surgery. The AR system showed promising results, compatible with otologic requirements, but also highlighted the need for extending the system to 3D for better visualization and ergonomics.

# 5 Acknowledgements

The authors would like to thank Oticon Medical, France for their financial support and NVIDIA for donating the TITAN X processor under their GPU grant program.